\documentclass[12pt]{iopart}
\usepackage{iopams}  
\usepackage{graphicx}

\begin{document}

\title{Metastable level properties of the excited configuration $4p^{6}4d^{8}4f$}

\author{R. Karpu\v{s}kien\.e, P. Bogdanovich, R. Kisielius}

\ead{Rasa.Karpuskiene@tfai.vu.lt}
\vspace{10pt}
\begin{indented}
\item[]December 22, 2014
\end{indented}

\begin{abstract}
Metastable levels in rhodium-like ions with the ground configuration 
$4p^{6}4d^{9}$ and the excited configurations $4p^{6}4d^{8}4f$ and
$4p^{5}4d^{10}$ are investigated. The {\sl ab initio} calculations of the level
energies, radiative multipole transition probabilities are performed in a 
quasirelativistic Hartree-Fock approximation employing an extensive 
configuration interaction based on quasirelativistic transformed radial 
orbitals. A systematic trends in behavior of calculated radiative lifetimes of 
the metastable levels are studied for the ions from $Z=60$ to $Z=92$. 
The significance of the radiative transitions of higher multipole order 
($M2$ and $E3$) for the calculated radiative lifetimes is demonstrated and 
discussed.
\end{abstract}

\noindent{\bf Keywords:\/} quasirelativistic approach, many-electron ions, 
metastable levels, radiative lifetimes\\

\noindent{\bf PACS:\/} {31.10.+Z, 31.15.ag, 32.70.Cs}\\

\maketitle

\section{Introduction}
\label{intro}

Our previous investigation of the $4p^{5}4d^{N+1}$ and $4p^{6}4d^{N-1}4f$ 
configuration metastable levels \cite{pc} has revealed that the ions of the
rhodium isoelectronic sequence have large number of metastable levels with 
comparatively large radiative lifetimes. Therefore, there is an evident need for 
a more detailed study. The relevance and significance of the magnetic quadrupole 
($M2$) and electric octupole ($E3$) transitions for such metastable levels has
been quite comprehensively described in \cite{pc}. That investigation of the 
significance of the $M2$ and $E3$ transitions involved large number of 
atoms of different ionization stages, therefore the calculations were performed 
without inclusion of correlation corrections. 

Metastable levels, which do not have the electric dipole decay channels, are 
essential in study of both astrophysical and laboratory plasmas 
\cite{ET2014, ET2012, PQ2010} or in analysis of experimental spectra of various 
ions \cite{MLQ2014, Biem2007}, including very important tungsten ions (see, 
e.g., \cite{PQ2010, SEY2012, Clem2010}). As it was concluded in \cite{ET2014},
the parameters of high multipole order radiative transitions and the radiative 
lifetimes of metastable levels were ``essential building blocks in 
radiative-collisional models simulating spectra''. Traditionally, the radiative
lifetimes for the excited metastable levels of the ground configuration are 
determined from the calculated electric quadrupole ($E2$) and magnetic dipole 
($M1$) transition probabilities, see \cite{ET2012, SEY2014, VJ2012, PQ2012}. 
These types of radiative transitions also can be important when the radiative 
lifetimes of the levels of excited configurations are calculated 
\cite{pc, MLQ2014, SEY2012}. Very recently the multiconfiguration 
Dirac-Hartree-Fock approach was applied to determine the $E1$, $M1$, $E2$, and 
$M2$ transition parameters for the ions up to $Z=36$ (see \cite{PJ2014, PR2014} 
and the references therein).

Many of the excited opposite-parity configuration levels are metastable ones due 
to a limited number of the $E1$-decay channels when the ground configuration has 
just few levels. In such a situation, the $E3$ and $M2$ radiative transitions 
become important decay channels - although those transitions traditionally are 
considered as weak and insignificant for the calculated radiative lifetimes 
\cite{MLQ2014, YI2010, Safr2006}. The relativistic many-body perturbation theory 
was employed to investigate the Fe$^{9+}$ ion properties in \cite{YI2010}. 
There the details describing the rise and decay of metastable levels are given 
and discussed. Transitions originating from the W$^{28+}$ ion metastable levels 
were observed in EBIT experiment \cite{MLQ2014}. It was concluded  after 
additional relativistic many-body perturbation theory (RMBPT) calculations that 
some of the excited configuration $4p^64d^94f$ levels have ``extremely long 
lifetimes''. The radiative transitions in Ni-like ions from $Z=30$ to $Z=100$ 
were studied in \cite{Safr2006} using RMBPT which includes the Breit interaction.

In our earlier work \cite{pc} we have demonstrated that the metastable levels
of the excited configurations also exist for the ions other than Ni-like 
\cite{Safr2006} or Pd-like ions \cite{MLQ2014} that have all closed shells in 
the ground configurations. Similar metastable levels can exist in the ions with 
the ground configurations, which have open 4d shells, e.g., in the Rh-like ions 
with the ground configuration $4p^64d^9$. Unfortunately, to our best knowledge, 
there are no theoretical or experimental works investigating the metastable 
level properties of those ions.

The aim of the present work is to demonstrate how the correlation corrections 
can change calculated radiative lifetimes of the metastable levels. Another 
issue, which has emerged during our previous investigation, is the location of 
the $4p^{6}4d^{8}5s$ and other excited configurations. The location in the 
energy spectrum of the excited configurations is important as the additional 
decay channels can emerge for some levels of the configuration $4p^{6}4d^{8}4f$.

Traditionally, the levels are considered as the metastable ones when the electric
dipole ($E1$) transitions from these levels are forbidden by the selection rules 
for the total orbital quantum number $J$. The metastable levels have two 
important channels of the radiative decay. One channel is comprised from the 
magnetic dipole and electric quadrupole transitions to the levels of the 
same parity, usually to the levels of the same configuration; nevertheless, the 
transitions among the levels of different configurations with the same parity 
are also allowed. The second channel constitutes the magnetic quadrupole  
and electric octupole transitions to the levels of the different-parity 
configurations. The strongest radiative transitions are those connecting the
excited configuration levels to the ground configuration levels. Nevertheless, 
some radiative transitions connecting the levels of opposite-parity excited 
configurations also can be relatively strong and significant in the calculation
of radiative lifetimes.

The object of the present investigation is the configuration $4p^{6}4d^{8}4f$ 
levels with high values of the total angular momentum $J$ fand their possible 
decay channels. We demonstrate how their relative positions in the energy 
spectra and their radiative properties change along the rhodium isoelectronic 
sequence.

The calculations were performed in the quasirelativistic (QR) Hartree-Fock 
approximation \cite {pbor06, pbor07}, using several different configuration 
interaction (CI) expansion sets. As the first step, the energy spectra were 
calculated for the ions from $Z=48$ to $Z=92$ adopting a small CI expansion, 
which consisted of few configurations for each parity. In this particular 
approximation (further we call it CI1), all electrons, both from configurations 
under study and the admixed ones, were described purely by solutions of the 
quasirelativistic equations \cite{pbor06, pbor07}. After the results of the 
energy spectra were analyzed, the subsequent investigation of the rhodium-like 
ions was limited to the ions starting from $Z=60$. Further, at the second step, 
the correlation corrections were included using the extensive CI wavefunction 
expansion. In this case, the electrons of the admixed configurations were 
described by the solutions of the quasirelativistic equations and by the 
quasirelativistic transformed radial orbitals \cite{pbor08}.

\section{Calculation method}
\label{method}

We use the quasirelativistic Hartree-Fock approximation in our \textit{ab initio} 
calculation of the ion level energies and radiative transition parameters, 
such as the line strengths $S$, weighted oscillator strengths $gf$, radiative 
transition probabilities $A$. The quasirelativistic calculations are performed 
in a way described in our previous studies \cite{pboras11, pbrk13}. 
At the start, the QR equations for the ground configuration $4p^{6}4d^{9}$ 
are solved. Then the QR equations for the $4f$ and $5l$ (up to $5g$) radial 
orbitals (RO) are solved in a frozen-core potential. The relativistic 
corrections are included in the Breit-Pauli approach specially adopted for the 
quasirelativistic radial orbitals \cite{pbor08}. We employ the same RO basis 
both for the even and odd configurations.

The correlation corrections are included by applying the CI method. In our basic 
approach, called CI1, the CI includes only the configurations under research. 
The even configurations $4p^{6}4d^{9}$, $4p^{6}4d^{8}5s$ and $4p^{6}4d^{7}5s^{2}$ 
and the odd ones $4p^{5}4d^{10}$, $4p^{6}4d^{8}5p$ and $4p^{6}4d^{8}4f$ make the 
smallest CI expansion sets used in the calculations. No admixed configurations 
are included in this approach. The results obtained in this approach (CI1) are 
used for the rough evaluation of the energy spectra of the configurations under 
investigation and the radiative transitions of different multipole orders.

The subsequent CI expansion is formed from the set of the admixed configurations 
generated by virtually exciting one or two electrons from the $4l$ and $5l$ 
shells of the adjusted configurations. The virtually excited electrons up to 
$5g$ in the admixed configurations are described by the quasirelativistic RO.
Although this RO basis is rather limited, the total amount of the admixed 
configurations, which can interact with the adjusted ground configuration, 
reaches 72 for $Z=60$. For the complex of the odd configurations $4p^{5}4d^{10}$ 
and $4p^{6}4d^{8}4f$, the number of possible admixed configurations adds to 146.
The strongly interacting configurations are selected by selection methods 
\cite{pbrk01, pbrkam05} applied in all our previously described 
calculations. The selection criterion \cite{pbor08} is $1 \cdot 10^{-4}$. 
This means that all configurations with the mean weight larger than 
$ 1 \cdot 10^{-4}$ are included into the CI expansion. There are 40 admixed even 
configurations and 31 odd one after the selection rules are applied for the ion 
$Z=60$. The admixed configurations produce very large amount of configuration 
state functions (CSF). This number is reduced by applying the CSF reduction 
methods described in \cite{pbrkam02}.

For each ion in the isoelectronic sequence, its own quasirelativistic RO basis 
is determined, the list of admixed configurations is generated by the same 
virtual excitations, and the admixed configurations are selected using the same 
criterion. Therefore the CI expansion differs for each particular ion and the 
number of admixed configurations decreases with increasing nuclear charge $Z$. 
For $Z=92$, the CI expansion is significantly smaller, and only 28 even 
configurations and 22 odd ones are selected. The results obtained by such a way 
are labeled as CI2.

In order to perform more accurate and reliable calculations in the 
multiconfiguration approximation, the basis of determined quasirelativistic RO 
is supplemented with the transformed radial orbitals (TRO) described by variable 
parameters \cite{pbor08, pboras11, pbrk13}. The TRO are determined 
for the radial orbitals having the principal quantum number $6 \leq n \leq 10$
and for all possible values of the orbital quantum number $l$, i.e. $l \leq 9$. 
The admixed configurations are generated by virtually exciting one or two 
electrons from the $4l$ and $5l$ shells of the adjusted configurations. 
The virtually excited electrons are described by the quasirelativistic RO 
($n \leq 5$ and $l \leq 4$) and by the quasirelativistic TRO for the remaining 
virtually excited electrons. 

Due to the larger basis of the RO, the number of admixed configurations is 
larger than that in the previous CI2 expansion. 83 even configurations and 73 
odd ones are selected for the ion $Z=60$ using the same selection criterion 
$1 \cdot 10^{-4}$. The number of admixed configurations in the CI expansion 
consistently decreases along the isoelectronic sequence, and, for $Z=92$, there 
are just 42 even and 39 odd admixed configurations. The results obtained by such 
technique are named as CI3. The comparison of the results, determined in the CI2 
and CI3 approaches, indicates that the level radiative lifetime values are quite 
close along the isoelectronic sequence, the difference does not exceed 10\%.


\Table{\label{tau}
The radiative lifetimes (in ns) of the W$^{29+}$ ion configuration
$4d^84f$ metastable levels. 
}
\hline	
\multicolumn{1}{l}{level}&
\multicolumn{1}{l}{CI1}&
\multicolumn{1}{l}{CI2}&
\multicolumn{1}{l}{CI3}&
\multicolumn{1}{l}{CI4}\\
\hline
$4d^8(^3F)4f\,\,^4I_{15/2}$ & 2.16E+07 & 1.67E+07 & 1.71E+07 & 1.70E+07 \\
$4d^8(^1G)4f\,\,^2K_{15/2}$ & 4.64E+04 & 4.73E+04 & 4.77E+04 & 4.80E+04 \\
$4d^8(^3F)4f\,\,^4I_{13/2}$ & 5.54E+15 & 2.52E+15 & 3.76E+15 & 4.27E+15 \\
$4d^8(^3F)4f\,\,^4H_{13/2}$ & 6.04E+06 & 6.54E+06 & 6.70E+06 & 6.75E+06 \\
$4d^8(^1G)4f\,\,^2K_{13/2}$ & 3.18E+04 & 3.13E+04 & 3.28E+04 & 3.30E+04 \\
$4d^8(^3F)4f\,\,^2I_{13/2}$ & 1.60E+04 & 1.67E+04 & 1.65E+04 & 1.66E+04 \\
$4d^8(^1G)4f\,\,^2I_{13/2}$ & 3.24E+04 & 3.37E+04 & 3.41E+04 & 3.44E+04 \\
$4d^8(^3F)4f\,\,^4H_{11/2}$ & 1.82E+07 & 1.97E+07 & 2.12E+07 & 2.18E+07 \\
$4d^8(^3F)4f\,\,^2H_{11/2}$ & 6.03E+05 & 6.27E+05 & 6.54E+05 & 6.56E+05 \\
$4d^8(^3F)4f\,\,^4G_{11/2}$ & 4.52E+05 & 4.94E+05 & 4.92E+05 & 5.01E+05 \\
$4d^8(^3F)4f\,\,^4I_{11/2}$ & 1.58E+04 & 1.60E+04 & 1.61E+04 & 1.62E+04 \\
$4d^8(^1D)4f\,\,^2H_{11/2}$ & 8.10E+03 & 8.31E+03 & 8.36E+03 & 8.40E+03 \\
$4d^8(^3F)4f\,\,^4F_{9/2} $ & 5.35E+04 & 4.72E+04 & 4.78E+04 & 4.81E+04 \\
$4d^8(^3F)4f\,\,^2G_{9/2} $ & 3.02E+04 & 2.53E+04 & 2.52E+04 & 2.51E+04 \\
$4d^8(^1D)4f\,\,^2H_{9/2} $ & 1.49E+06 & 1.64E+06 & 1.69E+06 & 1.72E+06 \\
$4d^8(^1D)4f\,\,^2G_{9/2} $ & 3.90E+05 & 4.49E+05 & 4.53E+05 & 4.64E+05 \\
$4d^8(^3F)4f\,\,^4I_{9/2} $ & 1.69E+04 & 1.67E+04 & 1.68E+04 & 1.68E+04 \\
$4d^8(^3P)4f\,\,^4F_{9/2} $ & 2.22E+04 & 2.26E+04 & 2.29E+04 & 2.31E+04 \\
\hline
\endTable

The calculations employing even the larger CI expansion have been performed for 
several selected ions in order to verify applied approach and to prove the 
convergence of the CI. These results were determined using the same large base 
of RO as in the CI3 approach, but the selection criterion $1 \cdot 10^{-5}$ was 
applied. In Table~\ref{tau} they are named as CI4. The number of admixed 
configurations in the CI4 approach reaches several hundreds for each parity. 
These calculations are very complex, whereas their results differ from the CI3 
data only slightly. The differences among of the lifetime values obtained 
employing different CI expansions are demonstrated in Table~\ref{tau}. 

We must underline that we use the $LS$-coupling scheme for the identification 
of all investigated levels of the rhodium isoelectronic sequence. Although this 
identification can vary along the sequence, we use a uniform identification 
according to that of the tungsten ion ($Z=74$) levels.

This table contains the radiative lifetime values  of the W$^{29+}$ ion 
metastable levels determined using different CI expansion sets. The levels 
discussed in the present work are included in Table~\ref{tau}. The averaged 
difference between the values determined in the CI1 approach and those from the 
CI2 approach is approximately $10\%$, the averaged difference between the values 
determined in the CI2 approach and the CI3 approach is approximately $6\%$.
Further, the averaged difference between the results from the CI3 and CI4 
approximations decreases to nearly $1\%$. So it is clear that larger CI 
expansions are unnecessary as the determined lifetimes of metastable levels 
change just very slightly. The exception is the level $4d^8(^3F)4f\, ^4I_{13/2}$
with very large radiative lifetime. It is obvious that the convergence test 
fails for this level. Here the radiative transition probabilities are strongly 
affected by the cancellation effects causing the lifetime value to change 
significantly.

Definitely, the values obtained employing the large expansion CI4 are more 
accurate and reliable, but, for a systematic study of the rhodium-like ion 
properties, the results of the CI3 approach are accurate enough. Furthermore, 
although the number of the admixed configurations in CI3 is not very large, 
the total number of CSF, due to the open $4d$ shells, is very large even after 
the reduction procedure. Consequently, the calculation consumes very large 
amount of computing resources. This must be considered as an important factor 
since we investigate an isoelectronic sequence which has large number of ions. 

As it is mentioned in the beginning of this section, our calculation method is 
discussed extensively in \cite{pbor08}. All two-electron interactions are 
included in the same way as it is done in the conventional Breit-Pauli 
approximation. For this purpose, the codes \cite{cff91a, cff91b, cff91c} are 
adapted for the quasirelativistic radial orbital peculiarities together with our 
own computer codes.

The level energies, radiative lifetimes and spectroscopic parameters of the 
$E1$, $M1$, $E2$, $M2$ and $E3$ transitions calculated using the CI3 approach 
are published in the database ADAMANT (http://www.adamant.tfai.vu.lt/database) 
for the ions from $Z=60$ to $Z=92$.

\section{Results and discussion}
\label{result}

The ground configuration of the rhodium-like ions is $4p^{6}4d^{9}$. It has two 
energy levels, $^{2}D_{3/2}$ and $^{2}D_{5/2}$. The excited configuration 
$4p^{5}4d^{10}$ is also quite a simple one and has only two levels, 
$^{2}P_{1/2}$ and $^{2}P_{3/2}$. These two excited levels have an usual 
radiative decay channel through the electric dipole ($E1$) transitions, 
therefore they are not metastable levels. Further in this work, we do not 
discuss them. The excited configuration $4p^{6}4d^{8}4f$ has quite a large 
number of the levels with high values of the total angular momentum $J$. 
These levels can not decay to the ground configuration by the $E1$ transition. 
Therefore their radiative lifetimes can be large if the excited levels with the 
opposite parity, e.g., belonging to the configuration $4p^{6}4d^{8}5s$, 
are located above them. The location of the excited configurations, including 
the $4p^{6}4d^{8}5s$, $4p^{6}4d^{7}5s^{2}$ and $4p^{6}4d^{8}5p$ is discussed
further in this section.

\subsection{Configuration locations}

\begin{figure}
\includegraphics [scale=2.48]{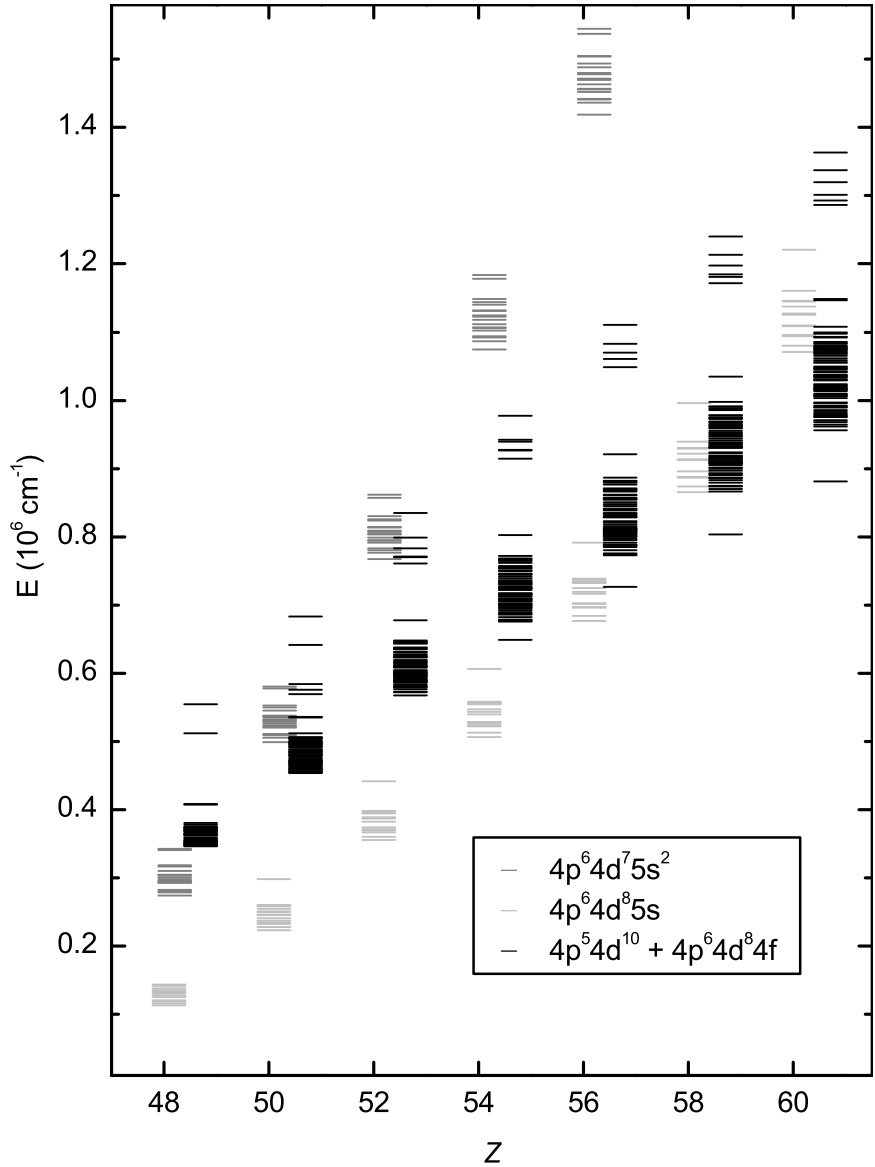} 
\caption{
\label{fig1} 
Energy levels (in $10^{6}$ cm$^{-1}$) of the even configurations $4p^{6}4d^{8}5s$, 
$4p^{6}4d^{7}5s^{2}$ and the odd-configuration complex $4p^{6}4d^{8}4f$ $+$ 
$4p^{5}4d^{10}$ relative to the lowest level $4p^{6}4d^{9}\; ^{2}D_{5/2}$ of each 
ion.
}
\end{figure}

At the start of our investigation, calculations in the CI1 approach along the 
isoelectronic sequence (from $Z=48$ to $Z=92$) are performed.  The level 
energies of the even configurations $4p^{6}4d^{9}$, $4p^{6}4d^{8}5s$, and 
$4p^{6}4d^{7}5s^{2}$ and the odd ones $4p^{5}4d^{10}$, $4p^{6}4d^{8}5p$, and 
$4p^{6}4d^{8}4f$ are determined by this rather simple way. As it is described in 
the previous section, only the interactions between configurations under 
research (adjusted configurations) are included. The CI1 results show how the 
configuration position changes with $Z$ increasing. A fraction of these results 
is presented in Fig.~\ref{fig1}. The levels of the odd configurations 
$4p^{5}4d^{10}$ and $4p^{6}4d^{8}4f$ and the even configurations 
$4p^{6}4d^{8}5s$ and $4p^{6}4d^{7}5s^{2}$ are plotted relative to the ground 
level $4p^{6}4d^{9}\; ^{2}D_{5/2}$.

When $Z=48$, the levels of the $4p^{6}4d^{8}4f$ configuration have additional
decay channels - not only those to the levels of the ground configuration. 
The decay channels to the levels of the $4p^{6}4d^{8}5s$ and 
$4p^{6}4d^{7}5s^{2}$ are open as these configurations are energetically close to 
the ground configuration and are lower than the $4p^{6}4d^{8}4f$ configuration
levels. As one can see from Fig.~\ref{fig1}, the energy difference between the 
$4p^{6}4d^{7}5s^{2}$ configuration and other configurations increases rapidly. 
Consequently, this configuration does not have any significance for the 
lifetimes of the $4p^{6}4d^{8}4f$ levels already at $Z=52$. The configuration 
$4p^{6}4d^{8}5s$ is located lower than the configuration $4p^{6}4d^{8}4f$ at 
$Z \leq 60$.

Using a strict definition, the $4p^{6}4d^{8}4f$ levels with $J \geq 9/2$ are 
not metastable ones in the case of such location of the excited configuration; 
only the levels with $J=15/2$ can be specified as metastable ones. The $E1$ 
transitions from the levels with $J = 15/2$ to the configurations 
$4p^{6}4d^{8}5s$ and $4p^{6}4d^{7}5s^{2}$ are forbidden by the selection rules 
for $J$. These transitions are allowed from the levels with $J=13/2$, but they 
appear only for the ion $Z=48$; moreover, they are weaker than the $M1$ and $E2$ 
transitions. For other ions, the levels with $J=13/2$ are metastable ones, 
because the levels of $4p^{6}4d^{7}5s^{2}$ are located higher.

\begin{figure}
\includegraphics [scale=2.48]{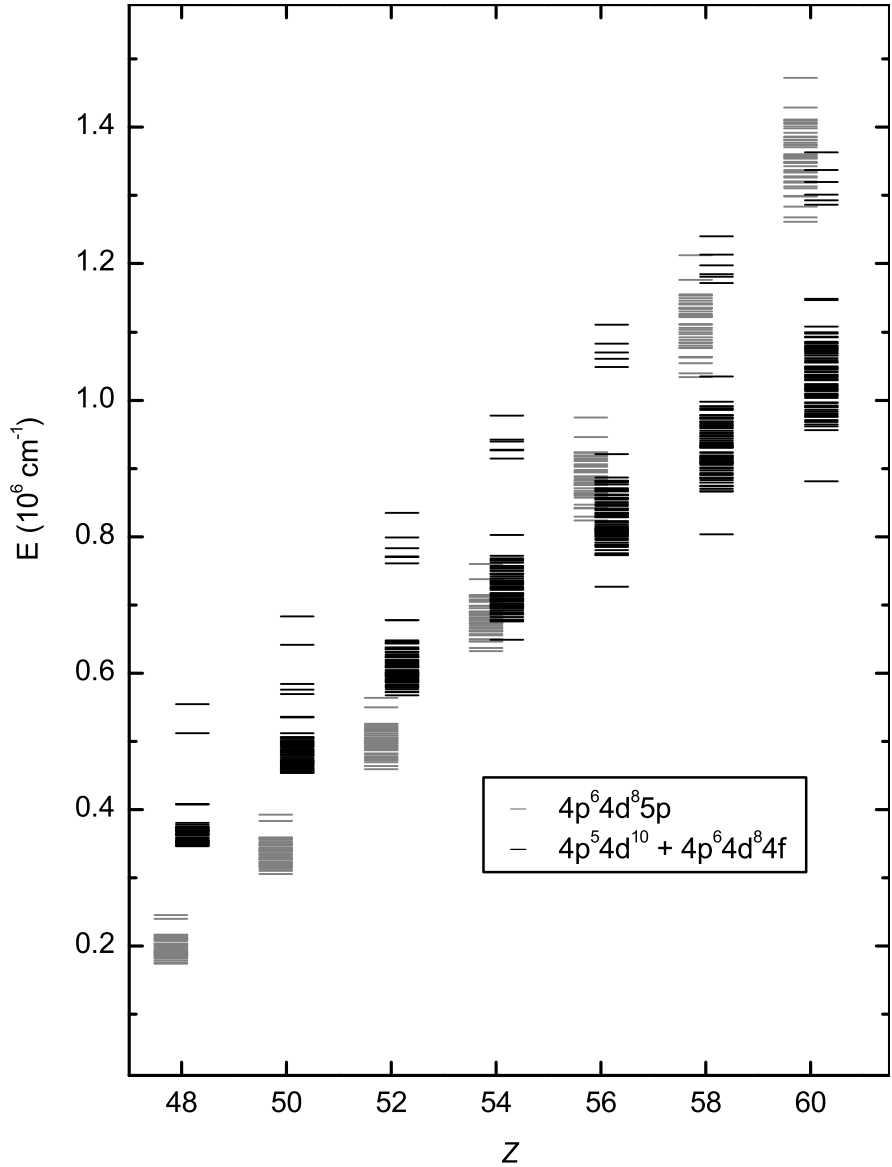} 
\caption{
\label{fig2} 
Energy levels (in $10^{6}$ cm$^{-1}$) of the odd configurations $4p^{6}4d^{8}5p$ 
and the odd complex $4p^{6}4d^{8}4f$ $+$ $4p^{5}4d^{10}$ relative to the lowest 
level $4p^{6}4d^{9}\; ^{2}D_{5/2}$ of each ion.
}
\end{figure}

Furthermore, the lifetimes of the levels with $J=13/2$ are reduced because of 
the $E2$ and $M1$ transitions to the configuration $4p^{6}4d^{8}5p$ of the same 
parity as the $4p^{6}4d^{8}4f$ one. The position of the $4p^{6}4d^{8}5p$ 
configuration in the energy spectra is presented in Fig.~\ref{fig2}. One can see 
from Fig.~\ref{fig2} that the $4p^{6}4d^{8}5p$ configuration is located in the 
same area as the $4p^{6}4d^{8}4f$ and $4p^{5}4d^{10}$ configurations for the 
ions $Z \leq 60$. Hence, for the reliable results, these configurations must be 
investigated as a single complex. Figures~\ref{fig1} and~\ref{fig2} demonstrate 
that the levels of the configuration $4p^{6}4d^{8}4f$ with high values of $J$ 
can be called the metastable ones only for the ions with $Z \geq 60$.

Figures \ref{fig1} and \ref{fig2} present the results of the simple CI1 approach. 
This approach is accurate enough for the basic calculation and for the initial 
assessment of the determined results. In order to determine more accurate values 
of the radiative transition parameters, the approaches described in the 
Sect.~\ref{method} with the larger CI expansions are applied for the ions 
from $Z=60$ up to $Z=92$. The results of the approaches CI2 and CI3 have been 
compared, and they indicate similar trends in properties of metastable levels. 
Nevertheless, we use the results of more reliable CI3 approach for the 
comprehensive study. The results CI4 are the most accurate data, but the CI4 
calculations were performed only for few ions, because of their complexity and 
large use of the computer resources.

\subsection{The lifetimes of $J=15/2$ levels}

\begin{figure}
\includegraphics[scale=2.48]{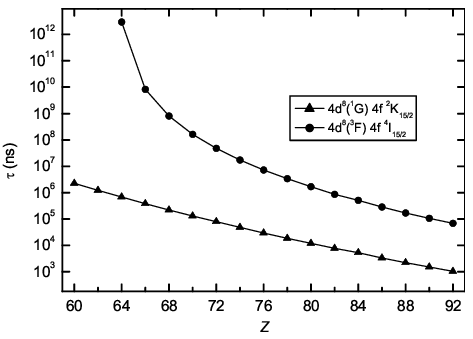} 
\caption{
\label{fig3} 
Radiative lifetimes (in ns) of the levels with $J=15/2$.
}
\end{figure}

The configuration $4p^{6}4d^{8}4f$ has two levels with $J=15/2$. The lower one 
is $4d^{8}(^{3}F)4f\; ^{4}I_{15/2}$ and the higher level is 
$4d^{8}(^{1}G)4f\; ^{2}K_{15/2}$. The radiative lifetimes of these levels are 
presented in Fig.~\ref{fig3} for the ions from $Z=60$ to $Z=92$.

\begin{figure}
\includegraphics[scale=2.48]{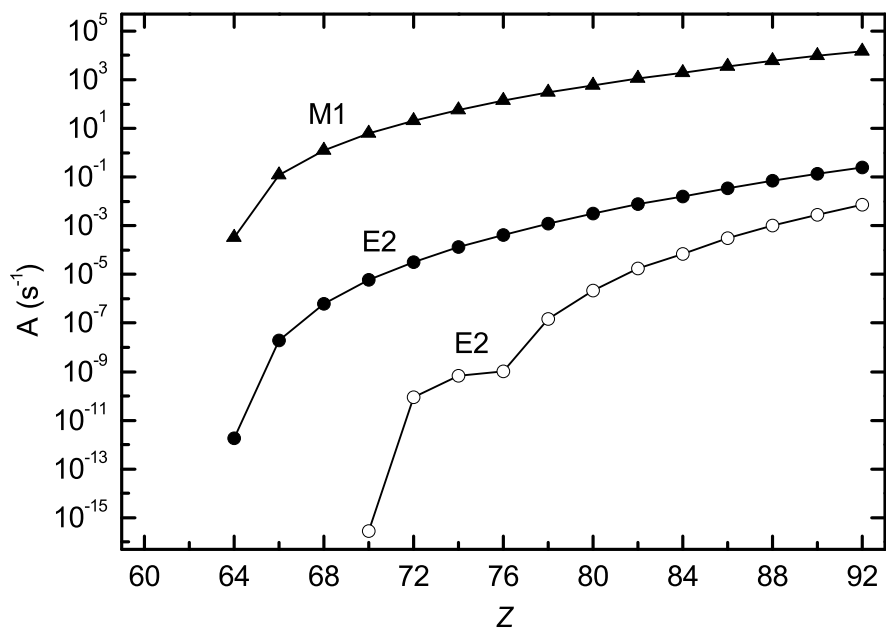} 
\caption{
\label{fig4} 
Transition probabilities (in s$^{-1}$) from the metastable level 
$4d^{8}(^{3}F)4f\; ^{4}I_{15/2}$. Black triangles and black circles denote 
transitions to the $4d^{8}(^{3}F)4f\; ^{4}I_{13/2}$, white circles - to the 
$4d^{8}(^{3}F)4f\; ^{4}H_{11/2}$.
}
\end{figure}

The radiative lifetime of the lower level is very large throughout the entire 
isoelectronic sequence due to its location in the energy spectra. When $Z=60$ 
and $Z=62$, this level does not have radiative transitions of the calculated 
multipole orders (in our case - up to $E3$) as there are no lower-lying levels 
with an appropriate $J$. With $Z$ increasing, the energy of the level 
$4d^{8}(^{3}F)4f\; ^{4}I_{15/2}$ increases. Moreover, it increases faster than 
the energy of some other levels. The transition probabilities for a decay of  
the level $4d^{8}(^{3}F)4f\; ^{4}I_{15/2}$ are presented in Fig.~\ref{fig4}.
The $E2$ transitions are denoted by circles, $M1$ transitions are denoted by 
triangles in this and in other figures, presenting the transition probabilities.

Another level with $J=15/2$, namely $4d^{8}(^{1}G)4f\; ^{2}K_{15/2}$, is 
located quite high in the energy spectra. Here the radiative lifetime is 
determined by some dozen $E2$ and $M1$ transitions with slowly increasing 
transition probability values. The strongest one is the $M1$ transition to the 
$4d^{8}(^{3}F)4f\; ^{4}I_{15/2}$ level. The transition probability value 
increases from $3.9\cdot 10^{2}$ s$^{-1}$ (for $Z=60$) to $8.4\cdot 10^{5}$ 
s$^{-1}$ (for $Z=92$). As it is seen from Table~\ref{tau}, for $Z=74$, the 
lifetime is $4.8\cdot 10^{4}$ s$^{-1}$.  In general, the $M1$ transitions 
determine the radiative lifetimes of this level for all investigated Rh-like 
ions.

\subsection{The lifetimes of $J=13/2$ levels}
\label{132}

\begin{figure}
\includegraphics [scale=2.48]{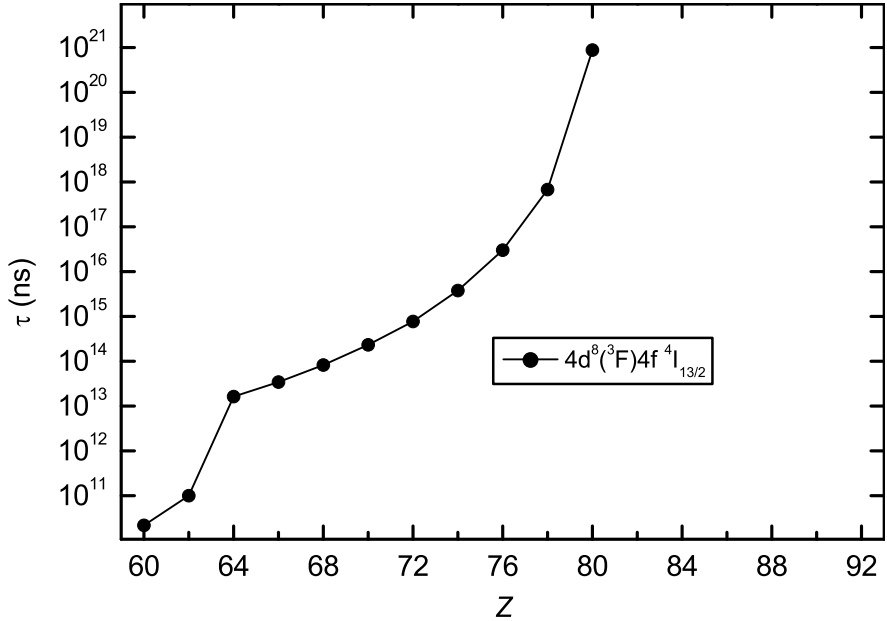} 
\caption{
\label{fig5} 
Radiative lifetimes (in ns) of the level with $J=13/2$.
}
\end{figure}

The configuration $4p^{6}4d^{8}4f$ has five levels with $J=13/2$. The radiative 
lifetimes of the lowest level $4d^{8}(^{3}F)4f\; ^{4}I_{13/2}$ are presented in 
Fig.~\ref{fig5}. The lifetimes of this level increase with the $Z$ increasing, 
because the location of this level in the energy spectrum is going down, i.e. 
the energy of this level relative to the ground level decreases unlike in the 
case of the lowest level with $J=15/2$, where the energy difference increases 
when the $Z$ increases.

The $4d^{8}(^{3}F)4f\; ^{4}I_{13/2}$ level has three decay channels open when 
$Z=60$ and $Z=62$. When $Z \geq 64$, only one $E2$ transition to the 
$4d^{8}(^{3}F)4f\; ^{4}I_{9/2}$ level is allowed. This transition determines the 
lifetimes of the level $4d^{8}(^{3}F)4f\; ^{4}I_{13/2}$ up to $Z=80$. For the 
higher $Z$, the decay channel to the $4d^{8}(^{3}F)4f\; ^{4}I_{9/2}$ is also 
closed. For the ions with $Z\geq82$, the $4d^{8}(^{3}F)4f ^{4}I_{13/2}$ level 
does not have a radiative decay channel, because the radiative transitions of 
all calculated multipole orders to the levels, which are lower than the 
investigated one, are forbidden by the selection rules for $J$. The lifetime of 
this level calculated for the tungsten ion is $3.8\cdot 10^{15}$ s$^{-1}$ (see 
Table~\ref{tau}), this is close to 49 days.

\begin{figure}
\includegraphics[scale=2.48]{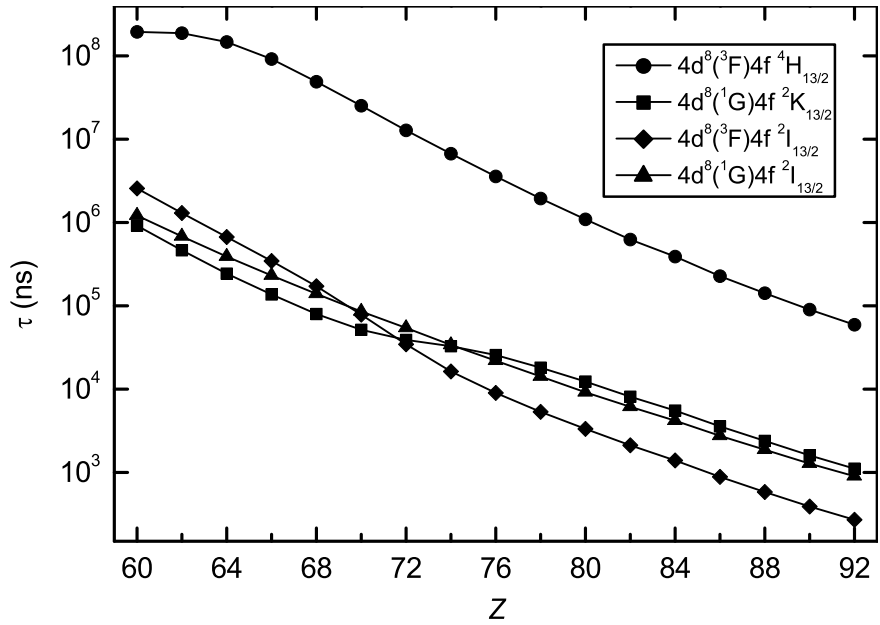} 
\caption{
\label{fig6} 
Radiative lifetimes (in ns) of four levels with $J=13/2$.
}
\end{figure}

\begin{figure}
\includegraphics[scale=2.48]{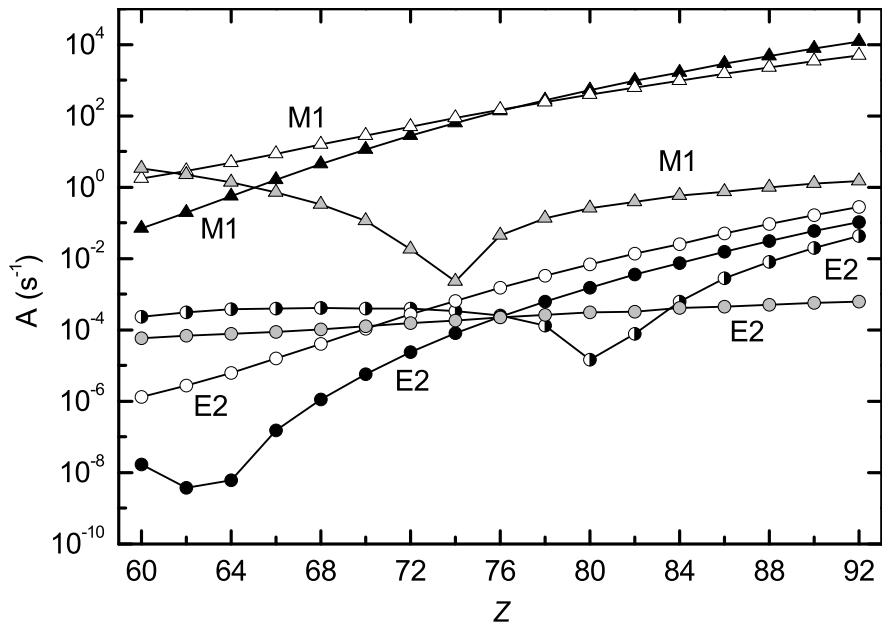} 
\caption{
\label{fig7} 
Transition probabilities (in s$^{-1}$) from the level 
$4d^{8}(^{3}F)4f\; ^{4}H_{13/2}$. Black triangles and circles denote transitions 
to the level $4d^{8}(^{3}F)4f\; ^{4}H_{11/2}$, white triangles and circles - to 
the level $4d^{8}(^{3}F)4f\; ^{4}I_{13/2}$, grey triangles and circles - to the 
$4d^{8}(^{3}F)4f\; ^{4}I_{15/2}$ level, black\&white circles - to the 
$4d^{8}(^{3}F)4f\; ^{4}F_{9/2}$ level.
}
\end{figure}

The lifetimes of other four levels with $J=13/2$ are presented in 
Fig.~\ref{fig6}. One can see from this figure that the radiative lifetimes 
consistently decrease going up along the sequence. The lifetimes of three levels 
are very similar in magnitude, but the lifetimes of the level 
$4d^{8}(^{3}F)4f\; ^{4}H_{13/2}$ are always noticeably (by two orders of 
magnitude) larger. This particular level is located relatively low in the energy 
spectra. Nevertheless, it is located higher compared to the level 
$4d^{8}(^{3}F)4f\; ^{4}I_{13/2}$. The level $4d^{8}(^{3}F)4f\; ^{4}H_{13/2}$ has 
seven allowed transitions to the lower levels, three of these being the $M1$ 
transitions and four the $E2$ ones. Their transition probabilities are presented 
in Fig.~\ref{fig7}. 

As one can see from Fig.~\ref{fig7}, the dependence of the transition 
probability$ A$ on a nuclear charge $Z$ has a marked minimum for some lines.
This is caused by cancellation effects. A radiative transition probability $A$ 
is proportional to a square of the transition operator matrix element. When this 
matrix element changes its sign, it's value becomes very close 0 at a particular 
value of $Z$. Consequently, the square of the radiative transition operator 
matrix element has a sharp minimum at this $Z$, causing such a peculiar shape of 
$A(Z)$. The similar shapes of $A$($Z$) can also occur in other isoelectronic 
sequences \cite{pc}. Very similar behavior of the radiative transition 
parameters proportional to a square of matrix element one can see in 
\cite{Safr2006}.

Another three levels with $J=13/2$ are located comparatively high in the energy 
spectra and have much more open decay channels. Furthermore, the strongest 
transition probabilities from these levels are from 10 to 100 times larger than 
those of the low-lying level $4d^{8}(^{3}F)4f\; ^{4}H_{13/2}$. The $M1$ 
transitions are dominant, as in Fig.~\ref{fig7}, and they determine the 
radiative lifetimes of the levels with $J=13/2$.

\subsection{The lifetimes of $J=11/2$ levels}

\begin{figure}
\includegraphics[scale=2.48]{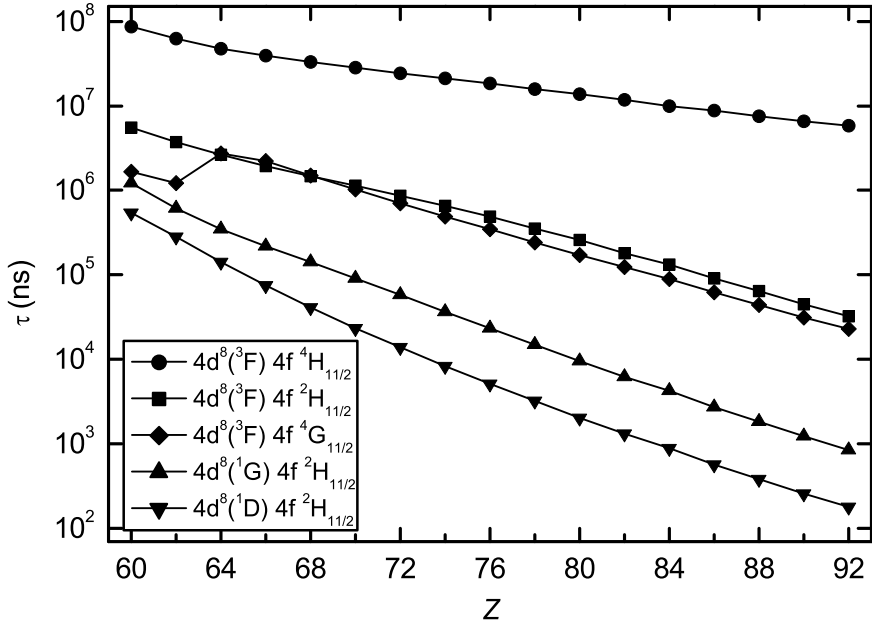} 
\caption{
\label{fig8} 
Radiative lifetimes (in ns) of five levels with $J=11/2$.
}
\end{figure}

\begin{figure}
\includegraphics[scale=2.48]{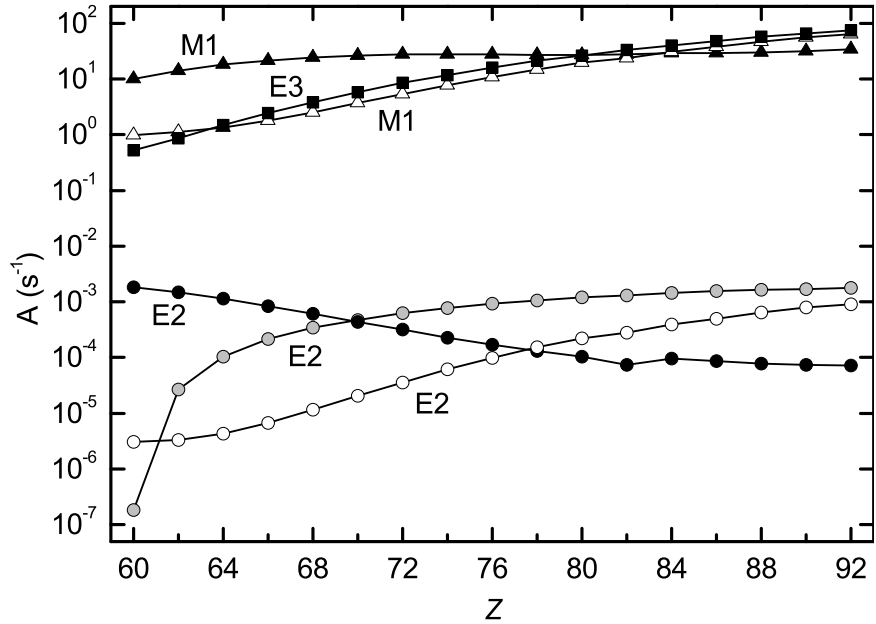} 
\caption{
\label{fig9} 
Transition probabilities (in s$^{-1}$) from the level 
$4d^{8}(^{3}F)4f\; ^{4}H_{11/2}$. Black triangles and circles denote transitions 
to the level $4d^{8}(^{3}F)4f\; ^{4}F_{9/2}$, white triangles and circles - to 
the level $4d^{8}(^{3}F)4f\; ^{4}I_{13/2}$, grey circles - to the 
level $4d^{8}(^{3}F)4f\; ^{4}F_{7/2}$, 
black squares - to the $4d^{9}\; ^{2}D_{5/2}$ level.
}
\end{figure}

The configuration $4p^{6}4d^{8}4f$ has nine levels with $J=11/2$. 
Figure~\ref{fig8} presents the radiative lifetimes of five levels, whereas the 
lifetimes of other four levels are in the region between the lowest two curves.
The level $4d^{8}(^{3}F)4f\; ^{4}H_{11/2}$ has the largest lifetime. The 
transition probability values of the radiative transitions from this level are 
presented in Fig.~\ref{fig9}. The $M1$ transition probabilities are 
significantly larger than the $E2$ transition probabilities. This feature is 
similar to properties of the levels with $J=13/2$.

Unlike the levels with $J \geq 13/2$, this level (as well as other levels with 
$J=11/2$) can decay to the ground configuration in accordance to the selection 
rules for $J$. This complementary decay channel is an electric octupole 
transition. The probabilities of the $E3$ transitions are marked by squares in 
Fig.~\ref{fig9} (and in the following plots, as they appear). The $E3$ 
transition probability for the $4d^{8}(^{3}F)4f\; ^{4}H_{11/2}$ level has the 
largest value at the top of the investigated isoelectronic sequence. That means 
that the $E3$ transition is the dominant one in determining the radiative 
lifetime. The inclusion of the $E3$ transition probability into the radiative 
lifetime calculation decreases the lifetime by factor of 1.8 at the top of the 
sequence. It is seen from Fig.~\ref{fig9} that the probabilities of two 
$M1$ transitions and one $E3$ transition determine the calculated radiative 
lifetime value of the $4d^{8}(^{3}F)4f\; ^{4}H_{11/2}$ level, because the 
probabilities of three $E2$ transitions are distinctly smaller. For the 
calculated lifetime of this level, the $E3$ transition is critical, especially 
at the top end of investigated sequence, where its value is largest. 
Other levels with $J=11/2$ also acquire an open decay channel via the $E3$ 
transition. Nonetheless, here the $E3$ transition probability is always somewhat 
smaller compared to several $M1$ transitions, although it is larger than the 
$E2$ transition probabilities. 
For these levels, the number of $M1$ transitions is large, but there is only one 
$E3$ transition, therefore it does not constitute a significant decay channel
as it does in the case of level $4d^{8}(^{3}F)4f\; ^{4}H_{11/2}$. 

Here we must explain that all calculated transition probabilities with the 
values larger than $1\cdot 10^{-18}$ s$^{-1}$ from the levels under 
consideration are included into determining of the radiative lifetime. It is 
important because only a limited number of transitions are allowed from the 
low-lying levels. If the level lies high in the energy spectra, the number of 
the $M1$ and $E2$ transitions is not so limited since a large number of the 
levels with appropriate $J$ values are located lower. Since there are only two 
levels of the ground configuration, the number of the $M2$ and $E3$ transitions 
always is one or two depending of the $J$ value of the initial level. Usually, 
the high-lying levels have relatively small radiative lifetimes as they have 
much more open decay channels to the levels of the same configuration.

The number of the transitions to the ground configuration is determined only 
by the selection rules. This number does not depend on the level location in the 
energy spectra.  The $E3$ transition probability values are similar in 
magnitude as the $M1$ probability values for a particular level. Nevertheless, 
the $E3$ transitions are rather insignificant decay channels for the high-lying 
levels as there is a large number of the $M1$ transitions allowed from these 
levels. Furthermore, the dominant $M1$ transition probabilities are larger for 
the high-lying levels, as in the case of levels with $J=13/2$.

\subsection{The lifetimes of $J=9/2$ levels}

\begin{figure}
\includegraphics [scale=2.48]{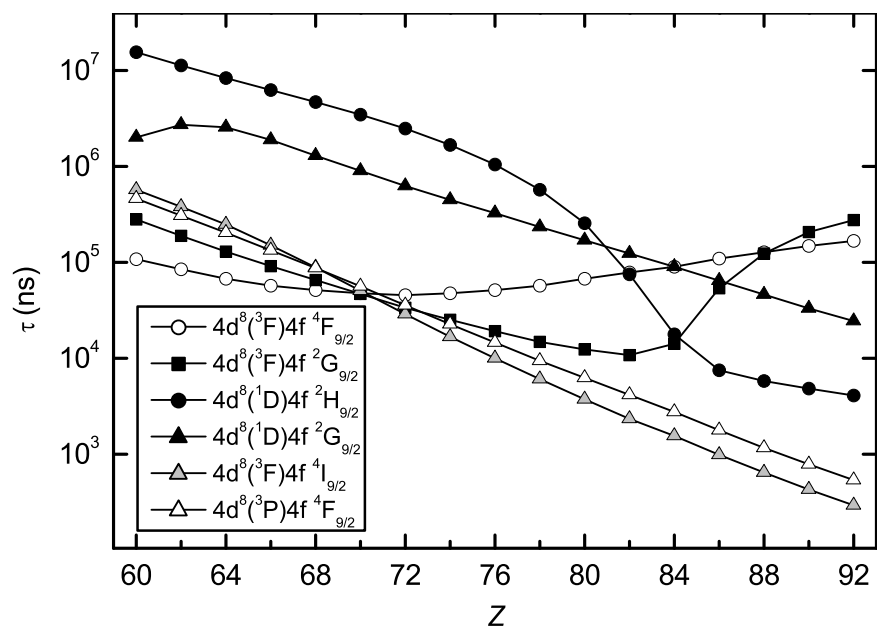} 
\caption{
\label{fig10} 
Radiative lifetimes (in ns) of six levels with $J=9/2$.
}
\end{figure}

The configuration $4p^{6}4d^{8}4f$ has thirteen levels with $J=9/2$. 
Figure~\ref{fig10} presents the radiative lifetimes only for six levels. 
The lifetimes of other seven levels are close to two curves, which present the 
lifetimes of levels $4d^{8}(^{3}F)4f\; ^{4}I_{9/2}$ and 
$4d^{8}(^{3}P)4f\; ^{4}F_{9/2}$ (white and gray triangles). The excluded levels 
are located higher than the presented ones, and their radiative lifetime 
dependencies on $Z$ are very similar to those of the latter two levels, both in 
a magnitude and in a shape. 

\begin{figure}
\includegraphics [scale=2.48]{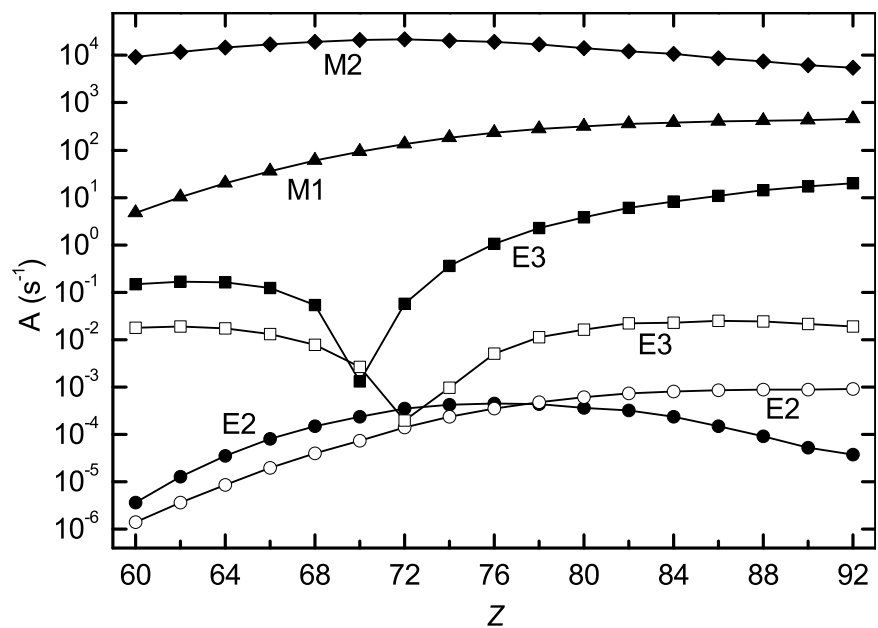} 
\caption{
\label{fig11} 
Transition probabilities (in s$^{-1}$) from the level 
$4d^{8}(^{3}F)4f\; ^{4}F_{9/2}$. Black triangles and circles denote transitions 
to the level $4d^{8}(^{3}F)4f\; ^{4}F_{7/2}$, white circles - to the level 
$4d^{8}(^{3}F)4f\; ^{4}F_{9/2}$, black diamonds and squares - to the level 
$4d^{9}\; ^{2}D_{5/2}$, white squares - to the level $4d^{9}\; ^{2}D_{3/2}$.
}
\end{figure}

The radiative lifetimes of the lowest level $4d^{8}(^{3}F)4f\; ^{4}F_{9/2}$ 
(white circles) are not the largest ones, unlikely to the previous ($J>9/2$) 
cases. These lifetimes are significantly smaller as there is an additional decay 
channel allowed by the selection rules. The transition probabilities for this 
level are presented in Fig.~\ref{fig11}. One can notice from Fig.~\ref{fig11} 
that the $M2$ transition (black diamonds) to the ground configuration is the 
strongest one. On top of that, two $E3$ transitions are allowed, and their 
transition probability values are larger than those of the $E2$ transitions.
These three radiative transitions ($M2$ and $E3$) to the ground configuration 
are very significant. At the lower end of the investigated sequence, their 
inclusion decreases the calculated radiative lifetime almost by 2000 times. 
Further, the influence of these transitions rapidly decreases, and the inclusion 
of them into the radiative lifetime calculations decreases the lifetime value 
approximately 13 times at $Z=92$. The reasons for a rapid decrease of the $E3$ 
transition values in Fig.~\ref{fig11} are explained in Sect.~\ref{132}. 
According to the selection rules for $J$, one $M2$ transition and two $E3$ 
transitions are allowed from this and all other levels with $J=9/2$. The number 
of the $M1$ and $E2$ transitions depends on the location of level examined. 

Next level with $J=9/2$ is the level $4d^{8}(^{3}F)4f\; ^{2}G_{9/2}$. 
Black squares denote the radiative lifetime of this level in Fig.~\ref{fig10}.
Here the $M2$ transition is the strongest one. But the $M2$ transition 
probability value starts to decrease slowly from $Z=84$. Several strong $M1$ 
transitions behave in a similar way. Such a drop of the transition probability 
values causes the radiative lifetime for this level to increase when 
$Z\geq84$. The $M2$ and $E3$ transitions are the most significant ones at 
$Z=64$ and $Z=66$. The inclusion of their probabilities into the the radiative 
lifetime calculations decrease the lifetime value by more than 450 times. 
At the top end of the sequence, this decrease drops to 3 times.

At the beginning of the sequence, the level $4d^{8}(^{1}D)4f\; ^{2}H_{9/2}$ has 
the largest radiative lifetime value (see Fig.~\ref{fig10}, black circles). 
The quite sharp change of the calculated lifetime for this level between $Z=80$ 
and $Z=86$ is caused by the rapid increase of the strongest $M2$ transition 
probability value together with that of several $M1$ transitions. 
Like for other previously discussed levels with $J=9/2$, there is one $M2$ 
transition, two $E3$ transitions and few $M1$ and $E2$ transitions. The $M2$ 
transition is the strongest one for $Z>70$, the $E3$ transitions are not so 
strong, but their transition probability values are larger than those of the 
$E2$ transitions. The $M2$ and $E3$ transitions affect the lifetime of 
$4d^{8}(^{1}D)4f\; ^{2}H_{9/2}$, but no so significantly as the lifetimes of 
previously discussed levels due to the amount of the $M1$ and $E2$ transitions 
allowed from this level.  

The radiative lifetimes of other levels with $J=9/2$, located higher in the
energy spectra than the discussed ones, are determined mainly by the numerous 
$M1$ transitions. The $E2$ transitions are also allowed, but they are much 
weaker than the $M1$ transitions. So here the influence of the $E3$ and $M2$ 
transitions is not so outstanding. Although these transitions cannot be 
discarded at all, when the nuclear charge $Z$ is between 60 and 70, these 
transitions decrease the calculated lifetime value up to 4 times. Usually, the 
decrease drops very quickly. For heavy ions ($Z>80$), the lifetime values change 
only by few percent.

\section{Summary}
\label{summ}

The configurations $4p^{6}4d^{9}$, $4p^{5}4d^{10}$, $4p^{6}4d^{8}5s$, 
$4p^{6}4d^{7}5s^{2}$, $4p^{6}4d^{8}5p$ and $4p^{6}4d^{8}4f$ of the Rh-like ions 
are investigated using the quasirelativistic approach with a small CI1 
expansion for the range from $Z=48$ up to $Z=92$. Calculation results 
demonstrate that the ions with $Z<60$ do not have metastable levels in the 
excited configurations, except for two levels of the the configuration 
$4p^{6}4d^{8}4f$ with the largest total angular momentum, $J=15/2$.

The levels  with $J=9/2$ of the $4p^{6}4d^{8}5s$ and $4p^{6}4d^{7}5s^{2}$ 
configurations are located relatively low in the energy spectra, and the $E1$ 
transitions from the $4p^{6}4d^{8}4f$ configuration levels with $J < 15/2$ are 
possible to the levels of these configurations. Furthermore, the decay channels 
through the $M2$ and $E3$ transitions to the  $4p^{6}4d^{8}5s$ and 
$4p^{6}4d^{7}5s^{2}$ are open. 

On the other hand, the decay channel through the $M1$ and $E2$ transitions is 
open for the $4p^{6}4d^{8}5p$ configuration. For the ions with $Z < 60$, the 
configuration $4p^{6}4d^{8}5p$ is close in energy to the $4p^{6}4d^{8}4f$ and 
$4p^{5}4d^{10}$. These energetically close and strongly interacting excited 
configurations must be investigated as a single configuration complex. 
The present investigation was devoted to the detailed investigation of the
configuration $4p^{6}4d^{8}4f$ metastable levels. 
The $4p^{6}4d^{8}4f$ configuration levels with $9/2 \leq J \leq 15/2$ are 
metastable ones for the highly-charged ions with $Z \geq 60$.  These ions are 
investigated thoroughly, using a large CI3 expansion within the basis of the 
quasirelativistic transformed radial orbitals.

The radiative lifetimes of the levels with $J=15/2$ and $J=13/2$ are determined 
by transitions to the levels with the same parity, i.e., the $M1$ and $E2$ 
transitions. It should be noted, that the $M1$ transition probability values for 
all investigated ions are much larger (more than 100 times) than the ones of the
$E2$ transition. The radiative lifetimes of the levels 
$4d^{8}(^{3}F)4f\; ^{4}I_{15/2}$ and $4d^{8}(^{3}F)4f\; ^{4}I_{13/2}$ are 
extremely large (more than 10 seconds) due to their position in the energy 
spectra. Although these two levels are relatively very low in the spectra, the 
level $4d^{8}(^{3}F)4f\; ^{4}I_{15/2}$ has three open radiative decay channels 
through rather weak transitions (one $M1$ and two $E2$). The level 
$4d^{8}(^{3}F)4f\; ^{4}I_{13/2}$ has only one decay channel through the $E2$ 
transition. 

Since the total number of possible $M1$ and $E2$ transitions depends on a 
particular level location, for the high-lying levels with $J=13/2$, as well as 
for other high-lying metastable levels, the $M1$ transitions primarily determine 
the calculated level radiative lifetime as their probability values are 
significantly larger compared to those of the $E2$ transitions. 

The lifetimes of the levels with $J=11/2$ and $J=9/2$ are also determined by 
the transitions to the levels of the same parity, but in this case, the $M2$ 
and $E3$ transitions to the ground configuration of the opposite parity are 
allowed. Since there are only two levels in the ground configuration, the number 
of $M2$ and $E3$ transitions is strictly defined. Only one $E3$ transition is 
possible from each level with $J=11/2$; two $E3$ transitions and one $M2$ 
transition is allowed from each level with $J=9/2$. The number of possible 
$E2$ and $M1$ transitions can vary depending on the location of the particular 
level in the energy spectra - when a particular level is located higher, the 
number of allowed $M1$ and $E2$ transitions becomes larger.

The $E3$ transition probability value is similar in magnitude with that of the 
$M1$ transitions and is larger than the transition probabilities of $E2$ for the 
levels with $J=11/2$. For the levels with $J=9/2$, the $M2$ transition 
probability has the largest value for most ions. The $E3$ transition probability 
values are smaller than those of the $M1$ transitions, but are larger than the 
$E2$ transition probabilities for these levels. 

The inclusion of $M2$ and $E3$ transitions into the radiative lifetime 
calculation can significantly change determined total radiative lifetime value. 
If the level position is comparatively high in the energy spectra, the $M2$ and 
the $E3$ transitions are not so significant, although they can not be neglected.
For the Rh-like ions, the decay of the metastable levels through the $M1$ and 
the $M2$ transitions is more significant compared to the decay through the $E3$ 
ones. The $E2$ transition probability values are the smallest ones for all 
metastable levels, except for the level $4d^{8}(^{3}F)4f\; ^{4}I_{13/2}$, which 
can decay only through the $E2$ transition.

\section*{Acknowledgment}

Current research is funded by the European Social Fund under the Global Grant 
measure, project VP1-3.1-{\v S}MM-07-K-02-013.


\section*{References}


\begin{thebibliography}{99}

\bibitem {pc}
R. Karpu\v{s}kien\.e, P. Bogdanovich and R. Kisielius, 
Significance of $M2$ and $E3$ transitions for $4p^{5}4d^{N+1}$ and 
$4p^{6}4d^{N-1}4f$ configuration metastable-level lifetimes, 
Phys. Rev. A {\bf 88}(2) 022519 (2013), 
http://dx.doi.org/10.1103/PhysRevA.88.022519

\bibitem{ET2014}
E. Tr\"{a}bert, 
$E1$-forbidden transition rates in ions of astrophysical interest, 
Phys. Scr. {\bf 89}(11) 114003 (2014), 
http://dx.doi.org/10.1088/0031-8949/89/11/114003

\bibitem{ET2012}
E. Tr\"{a}bert, 
Level lifetimes dominated by electric-dipole forbidden decay rates in the 
ground configuration of doubly charged rare gas ions (Ne$^{2+}$, Ar$^{2+}$, Kr$^{2+}$ and Xe$^{2+}$), 
Phys. Scr. {\bf 85}(4) 048101 (2012),  
http://dx.doi.org/10.1088/0031-8949/85/04/048101
 
\bibitem{PQ2010}
P. Quinet, V. Vinogradoff, P. Palmeri and \'{E}. Bi\'{e}mont, 
Radiative decay rates for W I, W II and W III allowed and forbidden transitions 
of interest for spectroscopic diagnostics in fusion plasmas", 
J. Phys. B: At. Mol. Opt. Phys. {\bf 43} 144003 (2010)
http://dx.doi.org/10.1088/0953-4075/43/14/144003
 
\bibitem{SEY2012}
S. Enzonga Yoca, P. Quinet and \'{E}. Bi\'{e}mont, 
Configuration interaction and radiative decay rates in trebly ionized tungsten 
(W IV), 
J. Phys. B: At. Mol. Opt. Phys. {\bf 45} 035001 (2012), 
http://dx.doi.org/10.1088/0953-4075/45/3/035001

\bibitem{MLQ2014}
M. L. Qiu, R. F. Zhao, X. L. Guo, Z. Z. Zhao, W. X. Li, S. Y. Du, J. Xiao, 
K. Yao, C. Y. Chen, R. Hutton and Y. Zou, 
Investigation of transitions between metastable levels of the first excited 
configuration of palladium-like tungsten", 
J. Phys. B: At. Mol. Opt. Phys. {\bf 47} 175002 (2014), 
http://dx.doi.org/10.1088/0953-4075/47/17/175002

\bibitem{Biem2007}
\'{E}. Bi\'{e}mont, A. Ellmann, P. Lundin, S. Mannervik, L.-O. Norlin, P. Palmeri, 
P. Quinet, D. Rostohar, P. Royen and P. Schef, 
Decay of metastable states in Nd II", 
Eur. Phys. J. D {\bf 41} 211-219 (2007), 
http://dx.doi.org/10.1140/epjd/e2006-00229-5

\bibitem{Clem2010}
J. Clementson, P. Beiersdorfer, and M. F. Gu,
X-ray spectroscopy of $E2$ and $M3$ transitions in Ni-like W" 
Phys. Rev. A {\bf 81} 012505 (2010), 
http://dx.doi.org/10.1103/PhysRevA.81.012505

\bibitem{SEY2014}
S. Enzonga Yoca, P. Quinet, 
Relativistic Hartree-Fock calculations of transition rates for allowed and 
forbidden lines in Nd IV, 
J. Phys. B: At. Mol. Opt. Phys. {\bf 47} 035002 (2014),  
http://dx.doi.org/10.1088/0953-4075/47/3/035002

\bibitem{VJ2012}
V. Jonauskas, G. Gaigalas, S. Ku\v{c}as, 
Relativistic calculations for $M1$-type transitions in $4d^{N}$ configurations of 
W$^{29+}$-W$^{37+}$ ions, 
Atomic Data and Nuclear Data Tables {\bf 98} 19-42 (2012), 
http://dx.doi.org/10.1016/j.adt.2011.08.001

\bibitem{PQ2012}
P. Quinet,
A theoretical survey of atomic structure and forbidden transitions in the 4pk 
and 4dk ground configurations of tungsten ions W$^{29+}$ through W$^{43+}$, 
J. Phys. B: At. Mol. Opt. Phys. {\bf 45} 025003 (2012), 
http://dx.doi.org/10.1088/0953-4075/45/2/025003

\bibitem{PJ2014}
P. J\"{o}nsson, P. Bengtsson, J. Ekman, S. Gustafsson, L.B. Karlsson, G. Gaigalas, 
C. Froese Fischer, D. Kato, I. Murakami, H.A. Sakaue, H. Hara, T. Watanabe, 
N. Nakamura and N. Yamamoto, 
Relativistic CI calculations of spectroscopic data for the 2p$^{6}$ and 2p$^{5}3l$ 
configurations in Ne-like ions between Mg III and Kr XXVII", 
Atomic Data and Nuclear Data Tables {\bf 100} 1-154 (2014),
http://dx.doi.org/10.1016/j.adt.2013.06.001

\bibitem{PR2014}
P. Rynkun, P. J\"{o}nsson, G. Gaigalas, C. Froese Fischer, 
Energies and $E1$, $M1$, $E2$, and $M2$ transition rates for states of the 
2s$^{2}$2p$^{3}$, 2s2p$^{4}$, and 2p$^{5}$ configurations in nitrogen-like 
ions between F III and Kr XXX", 
Atomic Data and Nuclear Data Tables {\bf 100} 315-402 (2014),
http://dx.doi.org/10.1016/j.adt.2013.05.003

\bibitem{YI2010}
Y. Ishikawa, J.A. Santana and E. Tr\"{a}bert, 
Relativistic multireference many-body perturbation theory for open-shell ions 
with multiple valence shell electrons: the transition rates and lifetimes of the 
excited levels in chlorinelike Fe X" 
J. Phys. B: At. Mol. Opt. Phys. {\bf 43} (2010) 074022 (2010), 
http://dx.doi.org/10.1088/0953-4075/43/7/074022

\bibitem{Safr2006}
U.I.Safronova, A.S.Safronova, S.M.Hamasha and P.Beiersdorfer, 
Relativistic many-body calculations of multipole ($E1$, $M1$, $E2$, $M2$, $E3$, 
and $M3$) transition wavelengths and rates between $3l^{-1}4l'$ excited and 
ground states in nickel-like ions" 
Atomic Data and Nuclear Data Tables {\bf 92} 47-104 (2006), 
http://dx.doi.org/10.1016/j.adt.2005.09.001

\bibitem {pbor06}
P. Bogdanovich and O. Rancova, 
Quasirelativistic Hartree-Fock equations consistent with Breit-Pauli approach, 
Phys. Rev. A {\bf 74}(5) 052501 (2006),
http://dx.doi.org/10.1103/PhysRevA.74.052501

\bibitem {pbor07}
P. Bogdanovich and O. Rancova, 
Adjustment of the quasirelativistic equations for p electrons, 
Phys. Rev. A {\bf 76}(1) 012507 (2007), 
http://dx.doi.org/10.1103/PhysRevA.76.012507

\bibitem {pbor08}
P. Bogdanovich and O. Rancova, 
Quasirelativistic approach for ab initio study of highly charged ions, 
Phys. Scr. {\bf 78}(4) 045301 (2008), 
http://dx.doi.org/10.1088/0031-8949/78/04/045301

\bibitem {pboras11}
P. Bogdanovich, O. Rancova and A. \v{S}tikonas, 
Quasirelativistic treatment of spectral characteristics of W$^{37+}$ 
Phys. Scr. {\bf 83}(6) 065302 (2011), http://dx.doi.org/10.1088/0031-8949/83/06/065302

\bibitem {pbrk13}
P. Bogdanovich and R. Kisielius, 
Theoretical energy level spectra and transition data for $4p^{6}4d^{2}$, 
$4p^{6}4d4f$, and $4p^{5}4d^{3}$ configurations of W$^{36+}$, 
At. Data Nucl. Data Tables {\bf 99}(5) 580-594 (2013), 
http://dx.doi.org/10.1016/j.adt.2012.11.001

\bibitem {pbrk01}
P. Bogdanovich and R. Karpu\v{s}kien\.e, 
Numerical methods of the preliminary evaluation of the role of admixed 
configurations in atomic calculations, 
Comp. Phys. Comm. {\bf 134}(3) 321-334 (2001), 
http://dx.doi.org/10.1016/S0010-4655(00)00214-9

\bibitem {pbrkam05}
P.Bogdanovich, R. Karpu\v{s}kien\.e and A. Momkauskait\.e, 
A program of generation and selection of configurations for the configuration 
interaction method in atomic calculations SELECTCONF, 
Comput. Phys. Commun. {\bf 172}(2) 133-143 (2005),         
http://dx.doi.org/10.1016/j.cpc.2005.06.006

\bibitem {pbrkam02}
P.Bogdanovich, R. Karpu\v{s}kien\.e and A. Momkauskait\.e, 
Some problems of calculation of energy spectra of complex atomic configurations, 
Comput. Phys. Commun. {\bf 143}(2) 174-180 (2002), 
http://dx.doi.org/10.1016/S0010-4655(01)00446-5

\bibitem{cff91a} 
A. Hibbert, R. Glass and C. Froese Fischer, 
A general program for computing angular integrals of the Breit-Pauli 
Hamiltonian, 
Comput. Phys. Commun. {\bf 64}(3) 445-472 (1991), 
http://dx.doi.org/10.1016/0010-4655(91)90138-B

\bibitem{cff91b} 
C. Froese Fischer, M.R. Godefroid and A. Hibbert, 
A program for performing angular integrations for transition operators, 
Comput. Phys. Commun. {\bf 64}(3) 486-500, (1991), 
http://dx.doi.org/10.1016/0010-4655(91)90140-G

\bibitem{cff91c} 
C. Froese Fischer and M.R. Godefroid, 
Programs for computing LS and LSJ transitions from MCHF wave functions, 
Comput. Phys. Commun. {\bf 64}(3) 501-519 (1991), 
http://dx.doi.org/10.1016/0010-4655(91)90141-7





\end{thebibliography}
\end{document}